\documentclass[a4paper,fleqn,usenatbib,useAMS]{mnras}

\usepackage{graphicx}
\usepackage{natbib}
\usepackage{color}
\usepackage{subfigure}
\usepackage{amsmath}

\def\hi{\relax \ifmmode {\mbox H\,{\scshape i}}\else H\,{\scshape i}\fi}

\def\hii{\relax \ifmmode {\mbox H\,{\scshape ii}}\else H\,{\scshape ii}\fi}

\def\nii{\relax \ifmmode {\mbox N\,{\scshape ii}}\else N\,{\scshape ii}\fi}

\def\oi{\relax \ifmmode {\mbox O\,{\scshape i}}\else O\,{\scshape i}\fi}

\def\oii{\relax \ifmmode {\mbox O\,{\scshape ii}}\else O\,{\scshape ii}\fi}

\def\oiii{\relax \ifmmode {\mbox O\,{\scshape iii}}\else O\,{\scshape iii}\fi}

\def\cii{\relax \ifmmode {\mbox C\,{\scshape ii}}\else C\,{\scshape ii}\fi}

\def\ciii{\relax \ifmmode {\mbox C\,{\scshape iii}}\else C\,{\scshape iii}\fi}

\def\civ{\relax \ifmmode {\mbox C\,{\scshape iv}}\else C\,{\scshape iv}\fi}

\def\hei{\relax \ifmmode {\mbox He\,{\scshape i}}\else He\,{\scshape i}\fi}

\def\heii{\relax \ifmmode {\mbox He\,{\scshape ii}}\else He\,{\scshape ii}\fi}

\def\mgii{\relax \ifmmode {\mbox Mg\,{\scshape ii}}\else Mg\,{\scshape ii}\fi}

\def\sii{\relax \ifmmode {\mbox S\,{\scshape ii}}\else S\,{\scshape ii}\fi}

\def\neiii{\relax \ifmmode {\mbox Ne\,{\scshape iii}}\else Ne\,{\scshape iii}\fi}

\def\ariv{\relax \ifmmode {\mbox Ar\,{\scshape iv}}\else Ar\,{\scshape iv}\fi}

\def\ni{\relax \ifmmode {\mbox N\,{\scshape i}}\else N\,{\scshape i}\fi}

\def\ariii{\relax \ifmmode {\mbox Ar\,{\scshape iii}}\else Ar\,{\scshape iii}\fi}

\def\caii{\relax \ifmmode {\mbox Ca\,{\scshape ii}}\else Ca\,{\scshape ii}\fi}

\title[]{Stellar content, planetary nebulae, and globular clusters of [KKS2000]04 (NGC1052-DF2)} 

\author[Ruiz-Lara]{T. Ruiz-Lara,$^{1, 2}$\thanks{E-mail: tomasruizlara@gmail.com (TRL)} I. Trujillo,$^{1, 2}$ M. A. Beasley,$^{1, 2}$ J. Falc\'on-Barroso,$^{1, 2}$ A. Vazdekis,$^{1, 2}$ \newauthor M. Filho,$^{3, 4}$ M. Monelli,$^{1, 2}$ J. Rom\'an,$^{1, 2}$ and J. S\'anchez Almeida$^{1, 2}$ \\
$^{1}$ Instituto de Astrof\'isica de Canarias, Calle V\'ia L\'actea s/n, E-38205 La Laguna, Tenerife, Spain \\
$^{2}$ Departamento de Astrof\'isica, Universidad de La Laguna, E-38200 La Laguna, Tenerife, Spain \\
$^{3}$ CENTRA-Ci\^{e}ncias, Science Faculty, University of Lisbon, Campo Grande, 1749-016 Lisbon, Portugal \\
$^{4}$ Department of Physics Engineering, Faculty of Engineering, University of Oporto, Rua Dr. Roberto Frias s/n, 4200-465 Oporto, \\ Portugal
}

\begin{document}

\date{Accepted 2019 April 30. Received 2019 April 24; in original form 2019 March 21}

\pagerange{\pageref{firstpage}--\pageref{lastpage}} \pubyear{2018}

\maketitle

\label{firstpage}

\begin{abstract}

[KKS2000]04 (NGC1052-DF2) has become a controversial and well-studied galaxy after the claims suggesting a lack of dark matter and the presence of an anomalously bright globular cluster (GC) system around it. A precise determination of its overall star formation history (SFH) as well as a better characterisation of its GC or planetary nebulae (PN) systems are crucial aspects to: i) understand its real nature, in particular placing it within the family of ultra diffuse galaxies; ii) shed light on its possible formation, evolution, and survival in the absence of dark matter. With this purpose we expand on the knowledge of [KKS2000]04 from the analysis of OSIRIS@GTC spectroscopic data. On the one hand, we claim the possible detection of two new PNe and confirm membership of 5 GCs. On the other hand, we find that the stars shaping [KKS2000]04 are intermediate-age to old (90\% of its stellar mass older than 5 Gyr, average age of 8.7 $\pm$ 0.7 Gyr) and metal-poor ([M/H] $\sim$ --1.18 $\pm$ 0.05), in general agreement with previous results. We do not find any clear hints of significant changes in its stellar content with radius. In addition, the possibility of [KKS2000]04 being a tidal dwarf galaxy with no dark matter is highly disfavoured.

\end{abstract}


\section{Introduction}

The agreement between numerical simulations and observations at large scales is astounding. However, discrepancies appear at smaller scales, where baryon physics has to be added to the simulations. On such scales, dwarf galaxies, the most abundant kind in the Universe, have proven crucial in our understanding of galaxy formation and evolution \citep[see][for a review]{2012RAA....12..917S}. Among them, a peculiar family of extended, yet of extremely low surface brightness galaxies, stands out \citep[e.g.][]{1984AJ.....89..919S, 1988AJ.....96.1520F, 1988ApJ...330..634I, 1991ApJ...376..404B, 1997AJ....114..635D, 2003AJ....125...66C, 2008MNRAS.383..247P, 2009MNRAS.393.1054P}. Although known since the 1980s, this family of objects \citep[currently known as ultra-diffuse galaxies, UDGs, ][]{2015ApJ...798L..45V} is receiving special attention as the emergence of new facilities, able to observe low surface brightness Universe without precendent, is enabling its massive detection \citep[][]{2015ApJ...798L..45V, 2015ApJ...807L...2K, 2016ApJS..225...11Y, 2016A&A...590A..20V, 2017MNRAS.468..703R}.

The current definition of UDGs, as introduced in \citet[][]{2015ApJ...798L..45V}, is based on the surface brightness of the object ($\mu_{\rm g}$(0)$\geq$24 mag arcsec$^{-2}$) and its effective radius (R$_{\rm e}$ $>$ 1.5 kpc), and thus, galaxies of different origins and characteristics can be included. While the growing consensus seem to suggest that UDGs are simply extended dwarf galaxies \citep[e.g.][]{2016MNRAS.459L..51A, 2016ApJ...819L..20B, 2016ApJ...830...23B,  2017MNRAS.466L...1D, 2017ApJ...836..191T, 2018ApJ...869...40S}, there are works claiming that there might be some embedded in huge dark matter haloes (like DF44 or DFX1, \citealt{2016ApJ...828L...6V, 2017ApJ...844L..11V, 2017MNRAS.464L.110Z}, but see also \citealt{2019arXiv190404838V}) or even lacking dark matter (like [KKS2000]04/NGC1052-DF2, \citealt[][]{2018Natur.555..629V}, see also \citealt[][]{2019ApJ...874L...5V}). [KKS2000]04\footnote{Although recently named as NGC1052-DF2 after its analysis in \citet[][]{2018Natur.555..629V}, along this paper we preferred to preserve the official name for this galaxy, [KKS2000]04 \citep[][]{2000A&AS..145..415K}.} is of particular interest because, in addition to the claim of lacking dark matter, it would also possess an unusual population of luminous globular clusters \citep[][]{2018ApJ...856L..30V}. However, a revision of the distance assumed to this galaxy has been proposed to solve both anomalies \citep[][]{2019MNRAS.tmp..733T}. A proper characterisation of its globular cluster (GC) system and stellar population properties is needed to place [KKS2000]04, a seemingly anomalous galaxy, within the realm of the regular UDGs.

UDGs are characterised by old, metal-poor stars \citep[][]{2017ApJ...838L..21K, 2018ApJ...859...37G, 2018ApJ...858...29P, 2018MNRAS.478.2034R, 2018MNRAS.479.4891F}, with the exception of some gas-rich, blue UDG progenitors found in lower density environments that might point towards an evolution scenario for UDGs \citep[][]{2017ApJ...836..191T, 2017MNRAS.468.4039R}. In this regard, recent photometric and spectroscopic studies suggest that [KKS2000]04 is indeed old and metal-poor \citep[][]{2018arXiv181207346F, 2019MNRAS.tmp..733T}, placing [KKS2000]04 as a normal UDG in terms of its stellar content. However, assuming a distance of 20 Mpc to this system, it still seems to hold a remarkably bright GC population \citep[][]{2018ApJ...856L..30V} unseen among UDGs until now.

As a consequence, a full characterisation of the GCs in [KKS2000]04 is of the utmost importance. On the one hand, completeness issues affecting the sample of GCs presented in \citet[][]{2018Natur.555..629V} and \citet[][]{2018ApJ...856L..30V} might be biasing the reported GC luminosity function. On the other hand, the number of GCs correlates with the halo mass, which is also true for UDGs \citep[][]{2019MNRAS.tmp..333P}. In addition, spectroscopic confirmation of as many of its GCs as possible would enable a more precise determination of the velocity dispersion of the system. As a consequence, a better knowledge on the GC population of [KKS2000]04 is needed to clarify all these aspects.

\citet[][]{2019MNRAS.tmp..733T} recently provided a new catalogue of GC candidates of [KKS2000]04. \citet[][]{2018arXiv181207345E}, making use of MUSE@VLT spectroscopic data and this catalogue, revised the velocity dispersion of the system by confirming one of the candidate GCs as well as adding to the analysis 3 newly-discovered planetary nebulae (PNe). In addition, they compute the stellar velocity dispersion of the main body of the galaxy through the analysis of an integrated spectrum. In both cases the velocity dispersion that they obtained ($\sigma_{\rm GC, PN}$~= ~10.6$^{+3.9}_{-2.3}$ km/s; $\sigma_{\star}$~=~10.8$^{-4.0}_{+3.2}$~km/s) was slightly higher but consistent within errors ($\sigma_{\star}$ consistent with values ranging from 5 to 13 km/s) than those originally suggested by \citealt[][]{2018Natur.555..629V} and later on by \citealt[][]{2019ApJ...874L..12D} ($\sigma_{\rm GC}$~=~7.8$^{+5.2}_{-2.2}$ km/s; $\sigma_{\star}$~=~8.5$^{+2.3}_{-3.1}$~km/s).

In this work we make use of OSIRIS@GTC long-slit spectroscopic data from [KKS2000]04 to analyse its stellar content, obtaining for the first time its star formation history (SFH) and assessing the possible existence of radial variations of the stellar properties in this system. We also expand on our knowledge of its GC system as well as search for new possible PNe. We describe the observations and the data reduction in Sect.~\ref{data}. The main results are described in Sect.~\ref{results}, while their main implications are discussed in Sect.~\ref{discussion}.

\section{Data reduction and analysis}
\label{data}

\begin{figure*}
\centering 
\includegraphics[width = 0.8\textwidth]{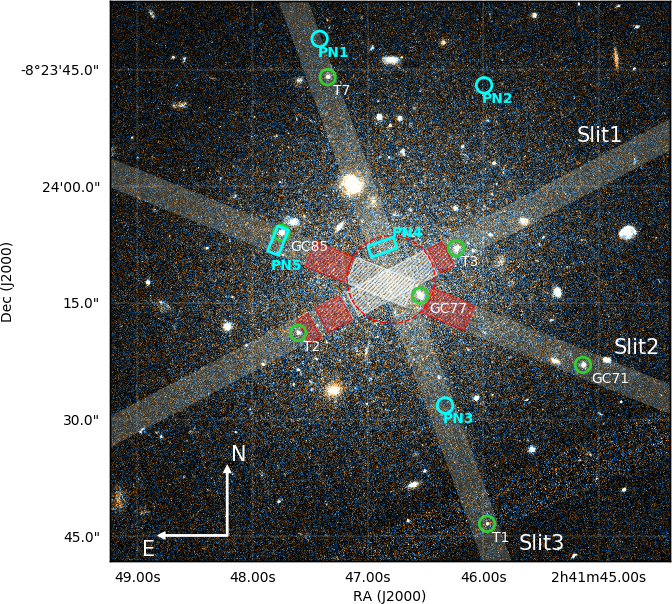} \\   
\caption{Colour composite image of [KKS2000]04 combining the F606W and F814W Hubble Space Telescope (HST) filters. The white shaded areas represent the 3 different OSIRIS slit configurations used in this work. Hatched areas delimit the area over which the integrated spectrum was computed: white for the integrated spectrum in the inner 1/4 R$_{\rm e}$ and red for the intermediate regions (up until $\sim$ 1/2 R$_{\rm e}$). For reference, the dashed, red circle encompasses the inner 1/4~R$_{\rm e}$ (R$_{\rm e}$~=~22.6''). The GCs analysed in this work are indicated as green circles, whereas the planetary nebulae detected are shown as cyan circles (those from Emsellem et al. 2018) or rectangles (new potential detections). GC39 is located towards the South-West outside of this zoom-in image, within {\it Slit3} (see e.g. van Dokkum et al. 2018 for the exact location of this GC).} 
\label{slit_conf} 
\end{figure*}

Deep spectroscopy is needed to study objects as faint as [KKS2000]04, with a central surface brightness of only $\mu$(V$_{\rm 606}$,0) = 24.4 mag arcsec$^{\rm -2}$. With this challenge in mind, we use the OSIRIS imager and spectrograph\footnote{For more information on the OSIRIS instrument we refer the reader to the instrument's webpage: \url{http://www.gtc.iac.es/instruments/osiris/}} \citep[][]{2000SPIE.4008..623C} mounted at the Gran Telescopio CANARIAS (GTC), in the Observatorio del Roque de los Muchachos. This instrument has been proven successful at recovering SFHs of UDGs before \citep[][]{2018MNRAS.478.2034R}. The OSIRIS observations analysed in this paper are part of the program \verb|GTC148-18B| (PI: I. Trujillo) and were carried out during dark, clear (seeing $\sim$ 1.2'') nights in November-December, 2018. Given the size and faintness of the target, we used OSIRIS in its long slit mode (effective length of 7.8') with a slit width of 3.5'' (maximizing the light entering the slit) and the R2000B grism. The combination of the selected grism and slit width results in a final spectral resolution of around 13.8~\AA, and a potential wavelength range coverage from 3950 to 5700~\AA. In total, 20 different exposures of 1400 seconds were taken in three different slit orientations as shown in Fig.~\ref{slit_conf}: {\it slit1} configuration, 6 exposures; {\it slit2} configuration, 6 exposures; and, {\it slit3} configuration, 8 exposures (totalling 10 hours on source). The slit orientations were chosen in order to maximize the number of already-analysed \citep[][]{2018Natur.555..629V} and candidate GCs discussed in \citet[][]{2019MNRAS.tmp..733T}. This strategy was especifically chosen to fulfil two main objectives. On the one hand, we wanted to collect as much light as possible to study the stellar content of the system. On the other hand, we were interested in the confirmation of membership of 4 new GCs proposed in \citet[][]{2019MNRAS.tmp..733T}\footnote{By the time the observations were performed none of the new GC candidates had any spectroscopic velocity measured.}, aware of the limitations of this observational strategy to obtain more precise radial velocities.

The reduction of all the individual exposures was performed independently using a pipeline specifically designed for dealing with OSIRIS-long slit data based on a set of \verb|PYTHON-IDL| routines. The pipeline follows all standard reduction steps such as bias subtraction, flat fielding, C-distortion correction, wavelength calibration, and cosmic rays removal \citep[L.A. Cosmic,][]{2001PASP..113.1420V}. Given the faintness of the object under analysis, special attention was paid to the sky subtraction. We use the {\it Kelson's sky subtraction algorithm} \citep[][]{2003PASP..115..688K} for such purpose. This technique relies on the knowledge of the CCD distortions and the curvature of the spectral features (previous reduction steps) to obtain a characteristic sky spectrum from carefully selected pixels with sky information. These sky pixels were carefully selected at the sides of [KKS2000]04 avoiding contamination from the galaxy as well as from other background or foreground sources. At this stage of the reduction, some artefacts such as ghosts (in ocasions as bright as the galaxy itself given its low surface brightness) and CCD imperfections were notable, especially in the spectral range around 4710--4760 \AA~that is avoided in the subsequent analysis. The outcome of this reduction scheme (i.e. 20 exposures fully-reduced 2D spectra) is treated differently for the different goals of this work.

\subsection{Point-like sources}
\label{point-sources}

The study of the GC membership (Sect.~\ref{GCs}) and the search for new PNe (Sect.~\ref{PNe}) were done after all exposures corresponding to each slit configuration were combined using \verb|IRAF| task \verb|imcombine| (\verb|median| combination with an \verb|avsigclip| rejection operation). The analysis of these three combined 2D spectra facilitates the location and extraction of the individual sources. Spectra of the GCs and GC candidates were extracted using {\tt APALL} in {\tt IRAF}. Background regions were defined by eye and fit with a 2nd-order polynomial in order (following the GC curvature in the CCD) to decontaminate  the GC spectra from the galaxy light that might be included because of the width of the slit. The spectra were extracted using optimal extraction based on the noise characteristics of the OSIRIS CCD. The GCs under analysis in this work are indicated in Fig.~\ref{slit_conf} as green circles. Following \citet[][]{2018arXiv181207345E} we have decided to identify the candidate GCs as ``T'' followed by a number given by the Table 2 in \citet[][]{2019MNRAS.tmp..733T}. For the already confirmed GCs we follow the notation introduced in \citet[][]{2018Natur.555..629V}. 

The detection of new PN candidates (line emitters in general) was made by visually inspecting the spectral region where the H$\beta$ and [\oiii] emission is expected (4850--5100 \AA), scanning the whole spatial extent covered by the three slits. For an easier and more robust detection, we clean those spectra from stellar emission by subtracting the mean spectrum of the adjacent pixels. The regions where the PN candidates found in this work are expected are highlighted in Fig.~\ref{slit_conf} as cyan rectangles (see Sect.~\ref{PNe}).

\subsection{[KKS2000]04 main body}
\label{main_body}

The brightest regions within [KKS2000]04 are combined together to obtain two integrated spectra from which the stellar content of the galaxy can be recovered. Because of the lower surface brightness of these regions, this time we prefer to work on the individual exposures rather than on the already combined exposures in each slit position. This approach allowed us to improve the statistics and perform a robust mean to obtain cleaner spectra, characterised by a higher S/N and less affected by the different artefacts of the OSIRIS CCD \citep[][]{2018A&A...617A..18R}. The two integrated spectra are representative of the inner 1/4 R$_{\rm e}$ of the galaxy and an intermediate region up to $\sim$ 1/2 R$_{\rm e}$, avoiding contamination from point sources (R$_{\rm e}$~=~22.6''). During the analysis of this dataset we found the presence of ghosts and CCD artefacts in the H$_{\rm \beta}$ and Mgb region for the {\it slit3} configuration exposures (spatially displaced with respect to the exposures corresponding to the other 2 configurations). For this reason, we decided to avoid those exposures in the extraction of the final two spectra. As a result, the effective time on source for the two integrated spectra is of 6 hours. In addition, the combination of source faintness and low sensitivity of the OSIRIS instrument in the blue part of the spectrum prevent us from using that particular region, delimiting the useful part of the spectrum from 4600 to 5400 \AA~(still a good wavelength range to carry out stellar population analysis).

\section{Results}
\label{results}

In this section we summarise the main results that we obtained regarding GC membership, search for new PNe, and the stellar content of [KKS2000]04.

\subsection{Revisiting the globular cluster membership}
\label{GCs}

The spectra corresponding to the [KKS2000]04 GCs (see Sect.~\ref{point-sources}) have been analysed using {\tt FXCOR} in {\tt IRAF} to determine their velocities and assess their real link with [KKS2000]04. The spectra were logarithmically rebinned, continuum-normalised with a low-order polynomial, and Fourier filtered to remove high-frequency noise and low-frequency continuum variations. We used three MILES model templates \citep[using the base models of][]{2010MNRAS.404.1639V} for the cross-correlation, with ages of 12 Gyr and metallicities of [M/H] = $-2.27$, $-1.79$ and $-1.49$ (with similar spectral characteristics as the ones we expect for the [KKS2000]04 GCs). We obtain velocities compatible with being associated to [KKS2000]04 for GC39, GC71, GC77 and GC85 \citep[][]{2018Natur.555..629V}, as well as T3 \citep[][]{2018arXiv181207345E}. In addition, we are able to rule out membership for T7 \citep[foreground star, see also][]{2018arXiv181207345E}. Unfortunately, we could neither confirm nor deny membership for T1 or T2 due to the low quality of its spectrum as a result of their extreme faintness (V$_{606}$ magnitude of 23.46 and 23.81, respectively). A quick comparison with the velocities obtained for the same objects in other works shows large offsets. However, given the observing strategy, not being specifically designed to obtain precise velocities for the GCs (wide slit and low spectral resolution), we expected large error bars and offsets with respect to previous works. Still, in the case of GC77, GC85, and T3, previous works report velocity values falling outside the range covered by our error determinations. This might suggest that uncertainties on the velocity measurements from the analysed data using {\tt FXCOR} are underestimated in some cases. Nevertheless, this observing strategy and methodology do allow us to confirm or rule out membership. Table~\ref{GCs_table} summarises our results for the 8 GC under analysis. 

\begin{table*}
{\normalsize
\centering
\begin{tabular}{llllllll}
\hline\hline
ID & RA & DEC & V$_{606}$ ($\pm$ 0.05) & E18 & vDK18 & V & Membership \\ 
 & (J2000) & (J2000) & (AB mag) & (km/s) &  (km/s) & (km/s) &  \\ 
$(1)$ & $(2)$ & $(3)$ & $(4)$ & $(5)$ & $(6)$ & $(7)$ & $(8)$ \\ \hline
GC39     & 40.43779 & --8.423583 & 22.38 & -- & 1818$^{+7}_{-7}$ & 1842 $\pm$ 69 & Yes \\
GC71     & 40.43807 & --8.406378 & 22.70 & -- & 1805$^{+6}_{-8}$ & 1868 $\pm$ 91 & Yes \\
GC77     & 40.44395 & --8.403900 & 22.03 & 1792.6$^{+4.4}_{-4.7}$ $^{+1.8}_{-11.6}$ & 1804$^{+6}_{-6}$ & 2079 $\pm$ 44 & Yes \\
GC85     & 40.44896 & --8.401659 & 22.42 & 1786.3$^{+4.3}_{-4.8}$ $^{+1.9}_{-6.1}$ & 1801$^{+5}_{-6}$ & 2024 $\pm$ 80 & Yes \\
T1$^a$   & 40.44153 & --8.412058 & 23.46 & -- & -- & --            & Unknown \\
T2$^a$   & 40.44836 & --8.405235 & 23.81 & -- & -- & --            & Unknown \\
T3$^b$   & 40.44265 & --8.402213 & 22.61 & 1788.7$^{+16.9}_{-10.9}$ $^{+16.3}_{-47.7}$ & -- & 2073 $\pm$ 105 & Yes \\
T7$^c$   & 40.44729 & --8.396096 & 22.21 & 22$^{+19}_{-16}$ & -- & 374 $\pm$ 113 & Foreground star \\
\hline
\end{tabular}

{\it a} The low quality of this spectrum hindered any velocity determination \\
{\it b} Spectroscopically confirmed as a [KKS2000]04 GC by Emsellem et al. 2018 \\
{\it c} Spectroscopically confirmed as a foreground star by Emsellem et al. 2018

\caption{Globular cluster membership. (1) Identifier for the GC from van Dokkum et al. 2018 and Trujillo et al. 2018; (2) Right Ascention (J2000); (3) Declination (J2000); (4) V$_{606}$ magnitude from Trujillo et al. 2018; (5), (6), and (7) Heliocentric-corrected velocity from Emsellem et al. 2018, van Dokkum et al. 2018, and this work, respectively.} 
\label{GCs_table}
}
\end{table*}

\subsection{On the search for new PNe}
\label{PNe}

\begin{figure}
\centering 
\includegraphics[width = 0.45\textwidth]{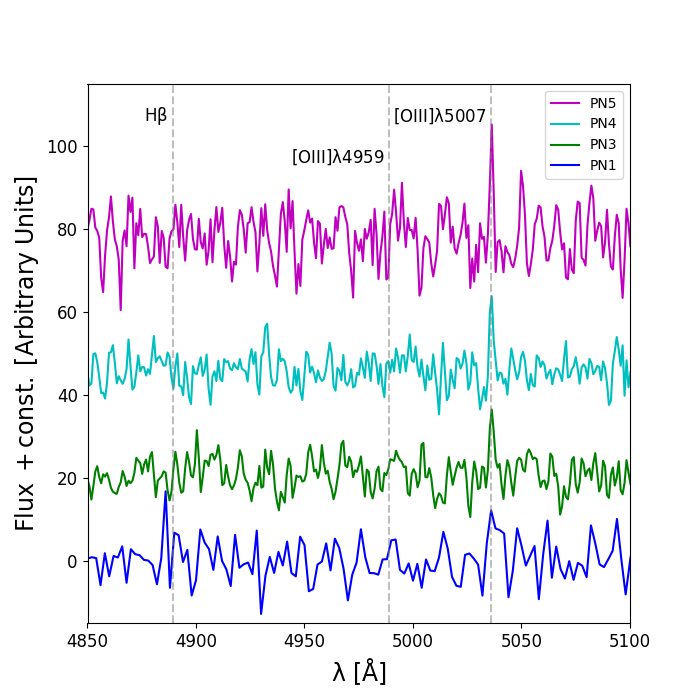} \\   
\caption{Spectra of the PNe candidates detected in this work. Offsets in the flux values have been applied to avoid overlapping between the spectra from the different (candidate and confirmed) PNe. The dashed vertical lines located at 4889, 4988 and 5036 \AA~correspond to the H$\beta$, [\oiii]$\lambda$4959 and [\oiii]$\lambda$5007 lines shifted according to the redshift of the galaxy.} 
\label{pn_spec} 
\end{figure}

Encouraged by the work by \citet[][]{2018arXiv181207346F} and \citet[][]{2018arXiv181207345E}, where the authors claim the detection of three PNe in [KKS2000]04, we examined the spectroscopic data presented in this work to search for new possible candidates. We confirm the presence of PN1 and PN3, the two PNe covered by this new dataset. In addition, we find the possible detection of another two line emitters at $\sim$ 5036 \AA, wavelength corresponding to the [\oiii]$\lambda$5007 line according to the redshift of the galaxy \citep[see][and this work]{2018Natur.555..629V, 2018arXiv181207346F, 2018arXiv181207345E, 2019ApJ...874L..12D}. The detection of this line is around 2~to~3~$\sigma$ above the noise level in both cases (see Fig.~\ref{pn_spec}). This detection, added to the theoretical intensity ratio between the two lines comprising the [\oiii] doublet of 2.98 \citep[][]{2000MNRAS.312..813S}, hampers the detection of the weakest line of the doublet ([\oiii]$\lambda$4959) with this set of data (within the noise level). We must bear in mind that the significance of our detections are highly minimised because of the fact that we are mixing together the information coming from a 3.5''-width area (slit width).

PNe are not the only point-like line emitters that can be found in galaxies. Supernova remnants or \hii~regions are other astrophysical objects that are compatible with being point-like sources in galaxies at distances of 13 Mpc or beyond \citep[as is the case of this system, see][]{2018Natur.555..629V, 2019MNRAS.tmp..733T}. In the blue, F([\oiii]$\lambda$5007)~$>$~F(H$\beta$) is usually regarded as a telltale for PNe. In particular, log($\rm \frac{[OIII]\lambda 5007}{H \beta}$) above 0.7 would ensure the PN nature of our detections \citep[see, e.g.][]{2010PASA...27..129F}. The detection of the [\oiii]$\lambda$5007 line in our PNe candidates (2~to~3~$\sigma$), together with the lack of detection of H$\beta$ (meaning signal around $\sigma$, see Fig.~\ref{pn_spec}) allow us to set a lower limit for log($\rm \frac{[OIII]\lambda 5007}{H \beta}$) of 0.5$^{+0.3}_{-0.1}$. Thus, although our detections seem to suggest that these point sources are PNe, we cannot discard them as possible \hii~regions. From herein, we will denote these point-like line emitters as PN candidates and not categorically as PN.

The reliable determination of the exact position of these PN candidates is also hindered by the observing configuration, namely the slit width. However, we can claim that they should be located within the rectangles given by these coordinates: PN4, [[40.4457, $-8.4025$], [40.4458, $-8.4021$], [40.4449, $-8.4018$], [40.4448, $-8.4022$]]; PN5, [[40.4491, $-8.4025$], [40.4495, $-8.4023$], [40.4491, $-8.4014$], [40.4487, $-8.4016$]]\footnote{Coordinates given by the location where emission is found and the width of the slit (3.5''). The format of the coordinates for each rectangle is as follows: [vertex${_1}$, vertex${_2}$, vertex${_3}$, vertex${_4}$] with vertex${_i}$ = [RA${_i}$, DEC${_i}$] in degrees.}. Visual inspection of HST images of this galaxy reveals the presence of point-like sources within both rectangles. Unfortunately, neither in the locations of the previously detected PNe (PN1, PN2, and PN3) nor in those of the new candidates (PN4 and PN5), clear photometric counterparts to these spectroscopic detections can be found in the available HST data. Indeed, some of the detected point-like sources might just be stars, hampering a proper identification of the mentioned PNe. For this reason, new sets of 2D spectroscopic and deeper photometric data are needed to fully assess the possible implications of these two new potential detections.

\subsection{[KKS2000]04 stellar populations}
\label{SFH}

We analyse the stellar content of the main body of [KKS2000]04 (see Sect.~\ref{main_body}) following an extensively tested and widely used methodology based on a series of full-spectral fitting codes \citep[e.g.][]{2011MNRAS.415..709S, 2014MNRAS.437.1534S, 2015MNRAS.446.2837S}. Although this technique is thoroughly explained in those works, here we summarise its main steps. First, we study the stellar kinematics shaping the observed spectrum and analyse the possible presence of gaseous emission using {\tt pPXF} \citep[penalized pixel fitting code; ][]{2004PASP..116..138C, 2011MNRAS.413..813C} and {\tt GANDALF} \citep[Gas AND Absorption Line Fitting;][]{2006MNRAS.366.1151S, 2006MNRAS.369..529F}, respectively. Once the stellar kinematics is recovered and the emission removed from the observed spectrum, we apply {\tt STECKMAP} \citep[STEllar Content and Kinematics via Maximum A Posteriori likelihood;][]{2006MNRAS.365...46O, 2006MNRAS.365...74O} to characterise its stellar populations. Because of the reported degeneracy between stellar metallicity and velocity dispersion when fitted simultaneously using full-spectral fitting techniques \citep[][]{2011MNRAS.415..709S}, we fix the stellar kinematics of the {\tt STECKMAP} run to the {\tt pPXF} values. Throughout all these steps we make use of the MILES\footnote{The models are publicly available at \url{http://miles.iac.es} and are based on the MILES empirical library \citep[][]{2006MNRAS.371..703S, 2011A&A...532A..95F}} models (base models, following the MILES stars chemical pattern) generated using the BaSTI \citep[][]{2004ApJ...612..168P} isochrones \citep[][]{2015MNRAS.449.1177V}. The recovered SFH and corresponding errors are computed following \citet[][]{2018A&A...617A..18R}, i.e. by combination of 12 different solutions scanning the whole {\tt STECKMAP} parameter space (errors from a set of 25 Monte Carlo simulations in each of the 12 individual solutions). Thus we minimise the impact of the {\tt STECKMAP} input parameters that might, somehow arbitrarily, affect the final solution. We should highlight that this technique has been proven successful at replicating the SFH recovered from the analysis of deep colour-magnitude diagrams from the Hubble Space Telescope \citep[][]{2015A&A...583A..60R}, in particular using OSIRIS long slit data of similar quality as the one studied here \citep[][]{2018A&A...617A..18R}.

This analysis allows us to obtain its stellar kinematics and characterise its stellar content. We obtain a systemic velocity of 1845 $\pm$ 46 km/s and 1784 $\pm$ 55 for the inner and intermediate regions of [KKS2000]04, in general agreement with the two previous determinations: \citet[][]{2019ApJ...874L..12D} found a value of 1805 $\pm$ 1.1 km/s, whereas \citet[][]{2018arXiv181207345E} state 1792$^{+1.4}_{-1.8}$ $^{+0.2}_{-1.3}$ km/s. Unfortunately, the low spectral resolution of our observational setup hampered the reliable determination of the stellar velocity dispersion of the system \citep[see][]{2018arXiv181207345E, 2019ApJ...874L..12D} as well as a more precise measurement of the velocities. Regarding the possible presence of ionised gas in the spectrum, {\tt GANDALF} was not able to detect any gaseous contribution to the observed spectrum with the required signal, suggesting that there is no ongoing star formation in [KKS2000]04. This is in agreement with the lack of \hi~gas found for this object \citep[][]{2004MNRAS.350.1195M, 2019MNRAS.482L..99C, 2019ApJ...871L..31S} as well as the absence of a prominent bright and blue main sequence in its colour-magnitude diagram \citep[][]{2019MNRAS.tmp..733T}. A visual inspection of the H$\beta$ region of the spectrum corresponding to the intermediate (1/4 R$_e$ $<$ R $<$ 1/2R$_e$) part of [KKS2000]04 might suggest the presence of some emission. However, the possible emission line detection is compatible with the noise and cannot be considered reliable, leading us to favour a technical origin behind such feature (problems with the CCD, ghosts, etc.). Assuming an upper limit for the H$\beta$ emission given by the noise of these data ($\sim 5\times 10^{-18}\, {\rm erg\,s^{-1}\,cm^{-2}\,\AA^{-1}}$), we can set an upper limit for the existing ionized gas mass in the range $2\times 10^{5-6} M_\odot$\footnote{We have used Eq.~(12) in \citet[][]{2017ApJ...834..181O} with the galaxy at 13 Mpc, the electron density in the range 10-100\, cm$^{-3}$, and an electron temperature of $10^4$ K.}, which is compatible with the mass expected from stellar evolution ejecta (around 1\%~of the stellar mass, i.e. $6\times 10^6 M_\odot$). 

\begin{figure}
\centering 
\includegraphics[width = 0.45\textwidth]{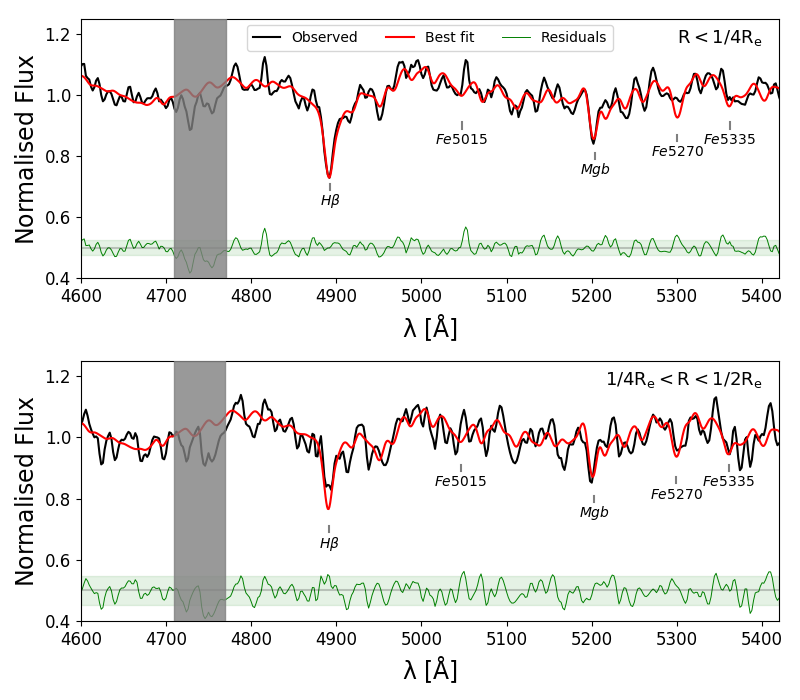} \\   
\caption{Inner (top panel) and intermediate-radius (bottom panel) observed, integrated spectra (black) for [KKS2000]04. The model spectra from {\tt STECKMAP} is shown in red together with the residuals in green. The shaded grey area depicts a spectral region not considered in the fit as it is an artefact in the OSIRIS CCD found in all slit configurations (see text for details). The shaded green area represents the expected residuals (noise) taking into account the S/N of the analysed spectra.} 
\label{gal_spec_SFH} 
\end{figure}

\begin{figure}
\centering 
\includegraphics[width = 0.45\textwidth]{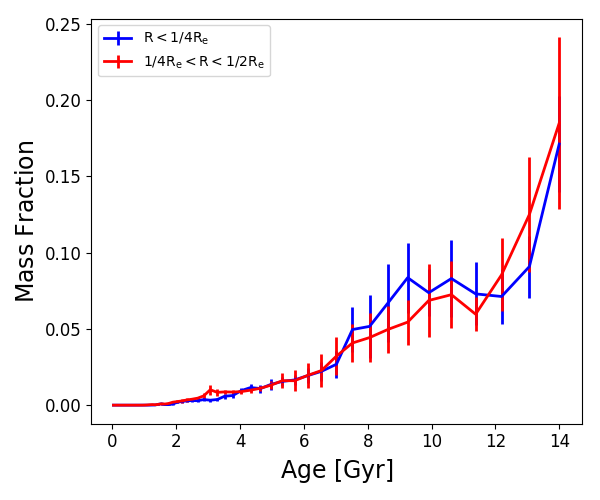} \\   
\caption{Recovered initial mass fraction as a function of age for the inner (blue, R $<$ 1/4R$_e$) and intermediate (red, 1/4 R$_e$ $<$ R $<$ 1/2R$_e$) parts of [KKS2000]04. Error bars are computed taking into account different tests using {\tt STECKMAP} scanning the whole input parameter space.} 
\label{gal_SFH} 
\end{figure}

Figure~\ref{gal_spec_SFH} shows the {\tt STECKMAP} fits (red) and the observed spectra (black) for the inner (R $<$ 1/4R$_e$) and intermediate (1/4 R$_e$ $<$ R $<$ 1/2R$_e$) parts of the galaxy. The fit is good given the quality of the spectrum and the systematic errors affecting the data reduction of such faint object (residuals always below 7\% and within the spectral noise given the S/N of the spectra, shaded green areas in the figure). The initial mass fractions as a function of age for the inner (blue) and intermediate regions (red) are displayed in Fig.~\ref{gal_SFH}. The bulk of the [KKS2000]04 population ($\sim$ 90\% in mass) is older than 5 Gyr, consistent with having little or no stars younger than 4-5 Gyr. Given the predominance of old ages (where present to initial mass transformations are almost constant according to the MILES models mass evolution), the difference between initial and present mass fractions is accounted by the errors. The average age is 8.7 $\pm$ 0.7 Gyr (8.6 $\pm$ 1.0 Gyr for the intermediate radius). The metallicity is basically constant with an average value of [M/H] $\sim$ --1.18 $\pm$ 0.05 dex for the inner parts and --1.1 $\pm$ 0.1 dex for the intermediate region. So, we can conclude that the SFH does not change over the analysed radial range, and thus, no radial change in age or metallicity is found. Table~\ref{galaxy_prop} summarises the main findings from the analysis of the two integrated spectra coming from the two regions under study.

\begin{table*}
{\normalsize
\centering
\begin{tabular}{ccccc}
\hline\hline
Region & S/N & V & LW Age (MW) & LW [M/H] (MW) \\ 
 & (pixel$^{-1}$) & (km/s) & (Gyr) &  (dex) \\ 
$(1)$ & $(2)$ & $(3)$ & $(4)$ & $(5)$ \\ \hline
 R $<$ 1/4R$_e$  &  43  & 1845 $\pm$ 46  &  8.7 $\pm$ 0.7 (9.8 $\pm$ 0.5) & --1.18 $\pm$ 0.05 (--1.20 $\pm$ 0.07) \\ 
 1/4R$_e$ $<$ R $<$ 1/2R$_e$  &  21  & 1784 $\pm$ 55  &  8.6 $\pm$ 1.0 (9.8 $\pm$ 0.9) & --1.1 $\pm$ 0.1 (--1.2 $\pm$ 0.1) \\ \hline
\end{tabular}
\caption{Parameters derived for the two analysed regions. (3) Heliocentric-corrected velocity computed using {\tt pPXF}; (4) and (5) average age and metallicity estimations (both, light and mass-weighted) from the SFH recovered combining all {\tt STECKMAP} tests.} 
\label{galaxy_prop}
}
\end{table*}

\section{Discussion}
\label{discussion}

[KKS2000]04 has received a lot of attention after the claims by \citet[][]{2018Natur.555..629V} that it lacks dark matter. Currently, most of the efforts focused on this galaxy concentrate on estimating its dynamical mass (using the GC system or the main body stellar velocity dispersion; e.g. \citealt[][]{2018Natur.555..629V, 2018arXiv181207346F, 2018arXiv181207345E, 2018ApJ...859L...5M, 2018RNAAS...2b..54V, 2018arXiv181207345E, 2019ApJ...874L..12D}) and a better determination of its distance \citep[e.g.][]{2019MNRAS.tmp..733T, 2018ApJ...864L..18V, 2018RNAAS...2c.146B, 2018arXiv181207346F}. Within this context, in this work we expand on the current knowledge on this galaxy with new spectroscopic data gathered with the GTC~telescope.

\subsection{[KKS2000]04 point-like sources}

A more robust characterisation of the GC system around this galaxy is one of the likely solutions to assess its apparently anomalous nature, improve the determination of the dynamical mass of the system as well as delimit the distance measurement. \citet[][]{2019MNRAS.tmp..733T}, after a careful inspection of HST images of [KKS2000]04, propose 8 new GC candidates of which one (T3) has already been confirmed and another one ruled out \citep[T7,][]{2018arXiv181207345E}. The addition of this new GC to the list of confirmed GCs [KKS2000]04 as well as three PNe allowed \citet[][]{2018arXiv181207345E} to revisit the velocity dispersion of the point-like sources, and thus, the dynamical mass of the system. They also obtain the stellar velocity dispersion of the [KKS2000]04 main body. In both cases the velocity dispersion that they obtain ($\sigma_{\rm GC, PN}$~= ~10.6$^{+3.9}_{-2.3}$ km/s; $\sigma_{\star}$~=~10.8$^{-4.0}_{+3.2}$~km/s) is slightly above those found by \citealt[][]{2018Natur.555..629V} (using the GC system, $\sigma_{\rm GC}$~=~7.8$^{+5.2}_{-2.2}$~km/s) and \citealt[][]{2019ApJ...874L..12D} (via the integrated spectrum of the main body, $\sigma_{\star}$ = 8.5$^{+2.3}_{-3.1}$~km/s). However, all velocity determinations are compatible within errors. In this work, we confirm the membership of 5 GCs (previously analysed in the literature) as part of this system. In addition, two new PN candidates are proposed, increasing the number of point sources in this system by 2. Unfortunately, the low spectral resolution of the data and the observational setup prevent us to measure velocities with the needed accuracy as to further constraint the dynamical mass of [KKS2000]04, but serve as starting point for follow-up observations.

The discovery of new PNe goes beyond the simple fact of adding new independent measurements to the velocity dispersion computation. \citet[][]{2018arXiv181207346F} use the luminonsity-specific PN number ($\alpha$ = N$_{\rm PN}$/L$_{\rm gal}$) to claim that the PN formation rate of [KKS2000]04 is similar to other galaxies. Following similar reasoning\footnote{Assuming that the detected PNe belong to the brightest 2.5 mag of the PNe luminosity function (PNLF), \citet[][]{2018arXiv181207346F} proposed that $\alpha$ = 10$\times$N$_{\rm PN, 2.5}$/L$_{\rm gal}$, with L$_{\rm gal}$ ranging between 6$\times$10$^{\rm 6}$ and 10$^{\rm 8}$ L$_{\odot}$.}, and given the number of assumptions made, the addition of two more PNe (still to be confirmed) would place the $\alpha$ parameter in the range between 4$\times$10$^{\rm -7}$ to 8$\times$10$^{\rm -6}$, still compatible with a typical metal-poor, old stellar system \citep[][]{2006MNRAS.368..877B}. If these two new PNe are confirmed as such\footnote{Quick inspection of the MUSE datacube analysed in \citet[][]{2018arXiv181207345E} and \citet[][]{2018arXiv181207346F} does not show clear evidence of the presence of these two candidate PNe (E. Emsellem and J. Fensch, private communication).}, the measurement of their luminosity would allow to improve the sampling of the PNLF, which could potentially result in another independent constraint to the distance to the galaxy. Unfortunately, the lack of spatial resolution in the OSIRIS dataset makes such detailed measurements impossible due to the inability to pinpoint the exact location of the sources. However, a simple comparison of the light that we detect for PN3 and the new PN4 (observed in the same slit configuration), allowed us to infer that the flux ratio between PN4 and PN3 is 0.83$^{+0.09}_{-0.12}$ (F$_{\rm PN4}$/F$_{\rm PN3}$). Thus, although PN4 could potentially help to better constrain the PNLF, its faintness hinders its use to further delimit the distance to [KKS2000]04 using the bright abrupt cut-off of the PNLF \citep[][]{2012Ap&SS.341..151C}.

\subsection{[KKS2000]04 stellar content}

\citet[][]{2018arXiv181207346F} provided the first spectroscopic determination of the age and metallicity of [KKS2000]04 (8.9 $\pm$ 1.5 Gyr and [M/H] $\sim$ --1.07 $\pm$ 0.12). These values are in perfect agreement with the ones that we derive for its inner (8.7 $\pm$ 0.7 Gyr and [M/H] $\sim$ --1.18 $\pm$ 0.05 dex) and intermediate-radius (8.6 $\pm$ 1.0 Gyr and [M/H] $\sim$ --1.1 $\pm$ 0.1 dex) parts. These age and metallicity determinations, together with Figure 14 in \citet[][]{2019MNRAS.tmp..733T}, question the validity of using the surface brightness fluctuations (SBF) method to estimate the distance to this galaxy. The predicted SBF ($\overline{M}_{\rm 814}$) for both, Padova00 \citep[][]{2000A&AS..141..371G} and BaSTI \citep[][]{2004ApJ...612..168P}, stellar tracks in this range of age and metallicity deviate significantly from the \citet[][]{2010ApJ...724..657B} relation between $\overline{M}_{\rm 814}$ and colour. Several other SBF calibrations can be found in the literature credible only within certain colour ranges \citep[][]{2018ApJ...856..126C, 2019arXiv190107575C} that should be used accordingly with the observed datasets. 

Apart from the average stellar age and metallicity of [KKS2000]04, which is also in excellent agreement with previous determinations for UDGs \citep[][]{2017ApJ...838L..21K, 2018ApJ...859...37G, 2018ApJ...858...29P, 2018MNRAS.478.2034R, 2018MNRAS.479.4891F}, the SFH recovered for this galaxy here (see Fig.~\ref{gal_SFH}) resembles those of the UDGs with known SFHs in the Coma cluster \citep[][]{2018A&A...617A..18R, 2018MNRAS.479.4891F}. In the case of the Coma UDGs, their structural and stellar population properties were interpreted as the outcome of the combination of internal processes and environmental effects within the cluster \citep[][]{2015MNRAS.452..937Y, 2016MNRAS.459L..51A, 2017MNRAS.466L...1D}. The similarity between the [KKS2000]04 SFH and those of Coma UDGs might suggest that similar processes were also important in [KKS2000]04, with the lack of younger stars pointing towards a halt of the star formation of this system long ago. In particular, ram pressure stripping and starvation might produce such an effect by: i) removing gas from this system or accelerating the use of the gas reservoir; ii) inhibiting the subsequent formation of stars; and iii) permitting the system to evolve passively until it acquires the current properties \citep[][]{2017MNRAS.468.4039R, 2017ApJ...836..191T}. However, we cannot discard the effect of internal processes such as supernova (SN) feedback in the evolution of this galaxy. Indeed, early SN explosions might have removed large amounts of gas, hampering star formation in [KKS2000]04. Nevertheless, extended SFHs might be  expected due to infalling of new gas and/or star formation from the remaining gas. In this sense, the extension in time of the recovered SFH might reinforce the claims favouring a closer distance to this galaxy (implying a lower stellar mass and thus, a lower impact of early SN feedback avoiding quenching).

\subsection{Could [KKS2000]04 be a tidal dwarf galaxy?}

The claimed lack of dark matter in [KKS2000]04 and its large peculiar radial velocity (regardless of the distance) could also support the hypothesis of this system being a tidal dwarf galaxy \citep[TDG,][]{1998A&A...333..813D, 2014MNRAS.440.1458D, 2015A&A...574A..93L, 2018Natur.555..629V, 2018arXiv181207346F}. However, on the one hand, large deviations from the expected radial velocity according to the Hubble flow have also been found in other UDGs \citep[][]{2019MNRAS.486..823R}, supporting the fact that recessional velocities are far from being a good indicator of distances for nearby galaxies \citep[see also][]{2019MNRAS.tmp..733T}. On the other hand, \citet[][]{2019MNRAS.tmp..733T} already analysed and discarded the possibility of [KKS2000]04 being a TDG based on the absence of gas in this system, its average stellar age and metallicity, dynamical mass, environment, and numerical simulations predictions. The recovered SFH for this system reinforces those claims. [KKS2000]04 completely stopped forming stars $\sim$ 2 Gyr ago, although uncertainties are consistent with being a quiescent galaxy since $\sim$ 4 to 6 Gyr ago (when $\sim$ 90\% of its mass was already in place). The large initial mass that [KKS2000]04 should have had in order to survive tidal stripping for that long \citep[][]{1998ApJ...498..143K} plays against a tidal origin for [KKS2000]04 if it was not embedded in a massive dark matter halo: a TDG with the [KKS2000]04 properties and no dark matter should have been disrupted long ago \citep[][]{2018MNRAS.474..580P}. Apart from that, the location of [KKS2000]04 in the stellar mass-metallicity plane is fully compatible with it being a dwarf galaxy \citep[][]{2018arXiv181207346F}, not falling in the TDG  region \citep[][]{2000AJ....120.1238D, 2003A&A...397..545W}. 

\citet[][]{2018ApJ...866L..11B} reported the discovery of two UDGs associated with tidal material. The authors claim that these findings favour scenarios where UDGs might be expanded dwarf galaxies or the result of collapsed tidal material (TDG origin). However, the later scenario was highly disfavoured as both systems lack in \hi~gas and do not follow the TDG mass-metallicity relation. In the particular case of [KKS2000]04, in addition to the previous arguments against its TDG nature, deep observations of this system presented in \citet[][]{2019A&A...624L...6M} do not show any signs of tidal debris around it, although some stellar streams were detected around NGC1052. One might argue that, although the stars in [KKS2000]04 might have formed at least $\sim$ 4 to 6 Gyr ago, the galaxy might have acquired the current form later on (transformation to a TDG). Nonetheless, the lack of tidal debris around it plays against this possibility.

Another aspect to take into account in this discussion is the homogeneous spatial distribution of the stellar populations in [KKS2000]04 (both, in terms of average values and the global SFH). This might suggest that their stellar populations are well-mixed spatially, hinting that the system is relaxed now. Such uniformity could have been acquired via mergers or tidal interactions during the first stages of the evolution of the system, homogenising its properties but not being able to disrupt it. The presence of an important amount of dark matter avoiding disruption is crucial in this scenario. In this sense, \citet[][]{2018arXiv181207345E} speculate with the possibility that, the discrepant velocities for some of their analysed GC (T3 and GC93) and the lack of consistency of their derived velocity field with that of a simple oblate axisymmetric system point towards tidal stirring or accretion events shaping [KKS2000]04. However, we cannot discard that the spatial uniformity of the stellar content in the inner-to-intermediate regions of this galaxy might be consequence of an homogeneous formation of the object as a whole or passive evolution given the stellar velocity dispersion of the system.

\vspace{0.8cm}

In conclusion, most of the evidence tend to suggest that [KKS2000]04 is an old, metal-poor, normal low surface brightness dwarf, that should have been disrupted long ago in the absence of dark matter. Thus, the fact that it has not been disrupted disfavours the tidal dwarf galaxy hypothesis. This galaxy also shows an homogeneous distribution of its stellar populations suggesting its nature as a virialised object, an homogeneous formation, or an important role of early tidal interactions. The information on point sources (GCs and PNe candidates) presented in this work will also help to expand our knowledge on this controversial system.

\section*{Acknowledgements}

We thank the anonymous referee for very useful comments. We would like to thank Chris Brook for very helpful suggestions and discussions and Peter Weilbacher for comments. This research has been mainly supported by the Spanish Ministry of Economy and Competitiveness (MINECO) under grant AYA2016-77237-C3-1-P. TRL thanks support from Ministerio de Ciencia, Innovaci\'on y Universidades via Juan de la Cierva - Formaci\'on grant (FJCI-2016-30342). TRL and MM also acknowledge support from grant AYA2017-89076-P from MINECO. IT acknowledges financial support from the European Union's Horizon 2020 research and innovation programme under Marie Sklodowska-Curie grant agreement No 721463 to the SUNDIAL ITN network. MAB acknowledges support from the Severo Ochoa Excellence Scheme (SEV-2015-0548). MEF gratefully acknowledges the financial support of the ``Funda\c{c}\~{a}o para a Ci\^{e}ncias e Tecnologia'' (FCT - Portugal), through the grant SFRH/BPD/107801/2015. JSA acknowledges support from grant AYA2016-79724-C4-2-P from MINECO.

This research makes use of python (\url{http://www.python.org}); Matplotlib \citep[][]{hunter2007}, a suite of open-source python modules that provide a framework for creating scientific plots; and Astropy, a community-developed core Python package for Astronomy \citep[][]{astropy2013}.

\bibliography{bibliography}

\bsp

\end{document}